\documentclass[twocolumn,showpacs,preprintnumbers,amsmath,amssymb,prb]{revtex4-1}
\usepackage{mathrsfs}
\usepackage{graphicx}
\usepackage{dcolumn}
\usepackage{bm}
\usepackage{amsmath}
\usepackage{amsfonts}

\begin{document}

\title{Universal Scaling of Quantum Anomalous Hall Plateau Transition}
\author{Jing Wang}
\author{Biao Lian}
\author{Shou-Cheng Zhang}
\affiliation{Department of Physics, McCullough Building, Stanford University, Stanford, California 94305-4045, USA}

\begin{abstract}
We study the critical properties of the quantum anomalous Hall (QAH) plateau transition in magnetic topological insulators.
We introduce a microscopic model for the plateau transition in QAH effect at the coercive field and then map it to the network
model of quantum percolation in the integer quantum Hall effect plateau transition. Generally, an intermediate plateau with
zero Hall conductance could occur at the coercive field. $\sigma_{xx}$ would have double peaks at the coercivity while $\rho_{xx}$
only has single peak. Remarkably, this theoretical prediction is already borne out in experiment. Universal scaling of the transport
coefficients $\rho_{xy}$ and $\rho_{xx}$ are predicted.
\end{abstract}

\date{\today}


\maketitle

\section{Introduction}

The recent discovery of QAH effect in a magnetic insulator has attracted considerable interest in this new state of quantum matter~\cite{qi2006,qi2008,liu2008,li2010,wang2011,yu2010,onoda2003,xiao2011,ruegg2011,chang2013b,wang2013a,wang2013b}.
In a QAH insulator, theoretically predicted in magnetic topological insulators (TIs)~\cite{qi2006,qi2008,liu2008,li2010,wang2011,yu2010}, the strong spin-orbit coupling and ferromagnetic (FM) ordering combine to give rise to an insulating state with a topologically nontrivial band structure characterized by a finite Chern number~\cite{thouless1982,haldane1988}. In a beautiful experiment, the QAH effect has been discovered in Cr-doped (Bi,Sb)$_2$Te$_3$ magnetic TI~\cite{chang2013b}, where at zero magnetic field, the gate-tuned Hall resistance $\rho_{xy}$ exhibits quantized plateau at values $\pm h/e^2$ while the longitudinal resistance $\rho_{xx}\rightarrow 0$. The plateau transition is of particular interest, in which $\rho_{xy}$ changes from one quantized value to another over a narrow interval of external magnetic field at the coercivity, and $\rho_{xx}$ exhibits peaks~\cite{chang2013b}. In this paper, we address the critical properties of the quantum phase transition between adjacent QAH phases, and some of the theoretical predictions are already confirmed in the QAH experiment~\cite{chang2013b}.

This issue is closely related to the integer quantum Hall effect (QHE) plateau transition~\cite{prange1990}. In a strong magnetic field $B$, a two-dimensional (2D) electron gas exhibits the QHE over a wide range of sample disorder. The plateau transition between different quantized value for $\rho_{xy}$ reflects delocalization transition in each Landau level (LL). This delocalization has shown to be a critical phenomena~\cite{kivelson1992,huckestein1995,sondhi1997,kramer2005}, where the localization length $\xi$ diverges as a power law $\xi\sim\left(B-B_c\right)^{-\nu}$ with a universal critical exponent $\nu$~\cite{pruisken1988,huckestein1990,huo1992}. Scaling behavior in transport coefficients has been observed as the zero-temperature critical point is approached, as a function of temperature $T$, sample size, and frequency, which yield the value $\nu\approx2.38$~\cite{li2009,koch1991b,engel1993}. Chalker and Coddington proposed a network model to describe the quantum percolation of 2D electrons in a strong magnetic field and a smooth random potential~\cite{chalker1988}. The semiclassical cyclotron orbits propagate along the equipotential lines of the disorder potential, and the tunneling processes occur whenever two orbits approach each other on a distance less than the cyclotron radius. Extensive numerical simulations~\cite{chalker1988,lee1993,slevin2009} show that the network model has a plateau transition with $\nu=2.4\pm0.2$, in excellent agreement with the experimental results.

\begin{figure}[b]
\begin{center}
\includegraphics[width=3.2in]{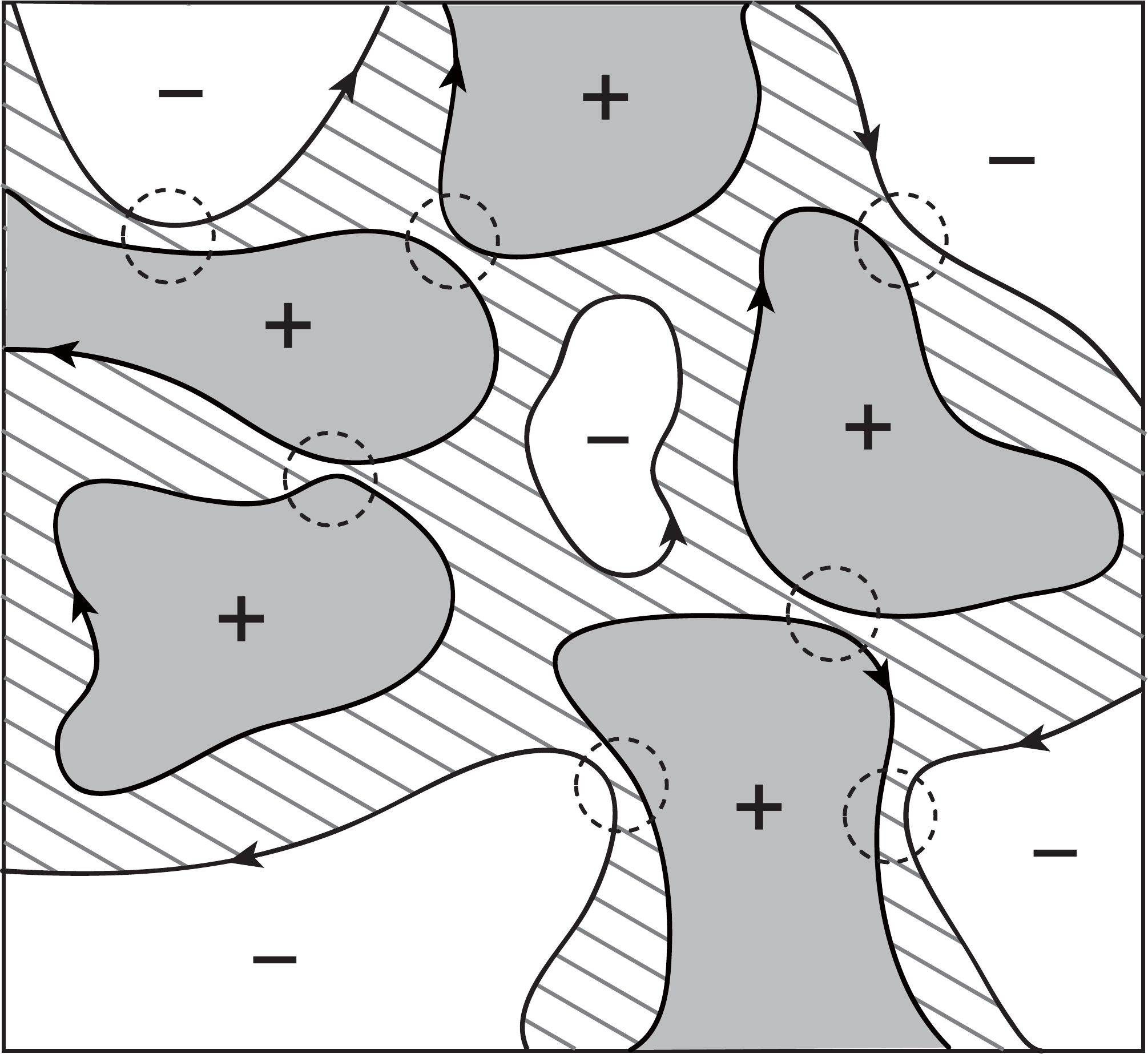}
\end{center}
\caption{Chiral edge states along domain walls at the coercivity in a magnetic TI. $+$ (grey region) and $-$ (white region) denotes the upward and downward magnetic domains with $|\Delta|>|m_0|$, respectively. The shadow region denotes $|\Delta|<|m_0|$. The arrowed lines are chiral states and correspond to the links in network model. The circles enclose the tunneling point between chiral states which correspond to the saddle points (nodes).}
\label{fig1}
\end{figure}

The magnetic TI studied in the QAH experiment~\cite{chang2013b} develop robust ferromagnetism at low temperature, possibly mediated by van Vleck mechanism~\cite{yu2010}. In the magnetized states, the magnetic domains of the material can be viewed as a single domain with up or down magnetization, and the system is in a QAH state with quantized $\rho_{xy}$ being $+h/e^2$ or $-h/e^2$. The magnetization reversal in this system leads to a quantum phase transition between two QAH states. At the coercive field, the magnetic domains are being switched from up to down \emph{randomly}, so many upward and downward domains coexist [marked as $+$ and $-$ in Fig.~\ref{fig1}]. At the boundary of each domain, there exists a chiral edge state~\cite{qi2008} with spatial decay length $\lambda$. Each edge state is characterized by a random phase change along the domain boundary. Tunneling between two edge states will occur whenever they are separated less than $\lambda$. Therefore, the QAH plateau transition at the coercivity in a magnetic TI is very much like the network model of the integer QHE plateau transition in the lowest LL. Although these two cases belong to quite different limits, the symmetries of the systems are common, i.e., the \emph{unitary} class without time-reversal nor spin-rotational symmetry~\cite{kramer2005}. One purpose of the present work is to propose a microscopic model for the QAH plateau transition, and establish its relation to the network model, so that the critical exponent obtained for the latter can be used for the former.

The organization of this paper is as follows. After this
introductory section, Sec. II describes the microscopic model
for the QAH plateau transition. Section III describes the mapping 
from the model for QAH plateau transition to the network model for 
the integer QHE transition. Section IV presents the
results and discussion on coercivity transition and experimental 
proposal in a magnetic TI. Section V concludes this paper. Some auxiliary materials are relegated into an Appendix.

\section{Model}

Now, we turn to the QAH state in 2D thin film of a magnetic TI with spontaneous FM order. The low-energy bands of this system consist of Dirac-type surface states only~\cite{qi2008,yu2010,wang2013a}, for the bulk states are always gapped. The generic form of the effective Hamiltonian is
\begin{eqnarray}\label{model0}
\widetilde{\mathcal{H}}_0(k_x,k_y)
   &=&v_F k_y\widetilde{\sigma}_1\otimes\widetilde{\tau}_3-v_F k_x\widetilde{\sigma}_2\otimes\widetilde{\tau}_3+\Delta\widetilde{\sigma}_3\otimes1
   \nonumber
   \\
   &&+m(k)1\otimes\widetilde{\tau}_2,
\end{eqnarray}
with the basis of $|t\uparrow\rangle$, $|t\downarrow\rangle$, $|b\uparrow\rangle$ and $|b\downarrow\rangle$, where $t$, $b$ denote the top and bottom surface states and $\uparrow$, $\downarrow$ represent the spin up and down states, respectively. $\widetilde{\sigma}_i$ and $\widetilde{\tau}_i$ ($i=1,2,3$) are Pauli matrices acting on spin and layer, respectively. $v_F$ is the Fermi velocity and we set $v_F\equiv1$. $\Delta$ is the exchange field along the $z$ axis introduced by the FM ordering. Here, $\Delta\propto \langle S\rangle$ with $\langle S\rangle$ the mean field expectation value of the local spin~\cite{yu2010}. The magnetization $M\propto\langle S\rangle_{\text{ave}}$ where $\langle S\rangle_{\text{ave}}$ is the spatial average of $\langle S\rangle$. $m(k)$ describes the tunneling effect between the top and bottom surface states. To the lowest order in $k$, $m(k)=m_0+m_1(k_x^2+k_y^2)$, and $\left|m_0\right|<\left|\Delta\right|$ guarantees the system is in the QAH state. For simplicity, the spatial inversion symmetry is assumed, which requires that $v_F$, $\Delta$ and effective $g$-factor take the same values for top and bottom surfaces.

In terms of the new basis $|+\uparrow\rangle$, $|-\downarrow\rangle$, $|+\downarrow\rangle$, $|-\uparrow\rangle$ with $|\pm\uparrow\rangle=(|t\uparrow\rangle\pm|b\uparrow\rangle)/\sqrt{2}$
and $|\pm\downarrow\rangle=(|t\downarrow\rangle\pm|b\downarrow\rangle)/\sqrt{2}$, the system is decoupled into two models with opposite chirality~\cite{wang2013a}
\begin{eqnarray}\label{model1}
\mathcal{H}_0(k_x,k_y) &=&
\begin{pmatrix}
\mathcal{H}_+(k) & 0\\
0 & \mathcal{H}_-(k)
\end{pmatrix},\\
\mathcal{H}_{\pm}(k) &=& k_y\tau_1\mp k_x\tau_2+\left(m(k)\pm\Delta\right)\tau_3
\end{eqnarray}
where $\tau_i$ are Pauli matrices. At half filling, $\mathcal{H}_{\pm}(k)$ have Chern number $\mp 1$ or $0$ depending on whether the Dirac mass is inverted ($m(k)\pm\Delta<0$) or not ($m(k)\pm\Delta>0$) at $\Gamma$ point. Thus the total Chern number of the system is
\begin{equation}\label{chern}
C = \begin{cases}
\Delta/|\Delta|, & \text{for} \ \ \left|\Delta\right|>\left|m_0\right|\\
0, & \text{for} \ \ \left|\Delta\right|<\left|m_0\right|
\end{cases}
\end{equation}
The Chern number changes by 1 at $\Delta=\pm m_0$. In the QAH state, the Hall conductance $\sigma_{xy}=Ce^2/h$ is in a quantized plateau and depends only on the sign of $\Delta$.

Magnetization reversal will change the sign of $M$, leading to the QAH plateau transition at $\Delta=\pm m_0$. Here we consider how the random magnetic domains at the coercivity will effect the QAH phase transition at $\Delta^*_1=m_0$ and $\Delta^*_2=-m_0$. In general, the disorder will generate spatially random perturbations to the pure Hamiltonian $\mathcal{H}_0$ in Eq.~(\ref{model1}). Specifically, at the coercivity, the system mainly has three types of randomness,
\begin{eqnarray}\label{randompotential}
\mathcal{H}_A &=& A_x(x,y)\tau_2\otimes\sigma_3-A_y(x,y)\tau_1\otimes1,
\nonumber
\\
\mathcal{H}_{\Delta} &=& \Delta(x,y)\tau_3\otimes\sigma_3,
\nonumber
\\
\mathcal{H}_{V} &=& V(x,y),
\end{eqnarray}
where $\sigma_3$ is Pauli matrix. $\vec{A}\equiv(A_x,A_y)$, $\Delta$, and $V$ are nonuniform and random in space, but constant in time. Thus they mix up the momenta but not the frequencies. $\mathcal{H}_A$ corresponds to a random vector potential, which comes from the gauge coupling ($\vec{k}\rightarrow\vec{k}-\vec{A}$) with the random magnetic field in the system. $\mathcal{H}_\Delta$ is the random exchange field along $z$ axis induced by the local spin in magnetic domains. $\mathcal{H}_V$ is the random scalar potential induced by impurities in the materials. Here the random exchange field within the $x$-$y$ plane is ignored, for effectively it only contributes to a negligible small random exchange field along $z$ axis at the transition point [see Appendix~\ref{randomperturbation}]. Obviously, $\mathcal{H}_A$ and $\mathcal{H}_\Delta$ break time-reversal symmetry, while $\mathcal{H}_V$ preserves time-reversal symmetry. To be concrete, at $\Delta=\pm m_0$, we will assume that all three random potentials are symmetrically distributed about zero mean. We also assume the interaction between the electrons can be neglected.

Here we mention that the model introduced above is very similar to the random Dirac model for the description of the integer QHE transition~\cite{fisher1985,ludwig1994}. The fixed point of the random Dirac model with all three different kinds of disorder is in a strong coupling regime, and is conjectured to be a generic integer QHE fixed point~\cite{ludwig1994}. This suggests the QAH plateau transition should have a similar critical behavior. However, the critical properties of the random Dirac model have not yet been accessible analytically. In order to get the critical exponents for QAH plateau transition, we construct a general mapping from the model for QAH transition to the network model.

\section{Mapping to network model}

Now, we consider $\mathcal{H}_+(k)$ in presence of disorders $\mathcal{H}_A$, $\mathcal{H}_\Delta$ and $\mathcal{H}_V$, which describes the phase transition from $C=+1$ to $C=0$ at $\Delta=-m_0$. In real space, the Hamiltonian has the form
\begin{equation}
\mathcal{H}_+=(-i\partial_y-A_y)\tau_1-(-i\partial_x-A_x)\tau_2+\delta\tau_3+V\ ,
\end{equation}
where $\delta(x,y)\equiv m_0+\Delta(x,y)$ is the Dirac mass. The $m_1$ term has been neglected, for it does not affect the plateau transition. For convenience, we make a unitary transformation $\widetilde{\mathcal{H}}_+\equiv G\mathcal{H}_+G^\dag$, and obtain
\begin{equation}
\widetilde{\mathcal{H}}_+ = (-i\partial_x-A_x)\tau_3-(-i\partial_y-A_y)\tau_1-\delta\tau_2+V,
\end{equation}
with $G=(\tau_2-\tau_3)/\sqrt{2}$. In the low-energy limit, the unitary evolution operator in a unit time for $\widetilde{\mathcal{H}}_+$ is
\begin{equation}\label{evolution}
\mathcal{U}=e^{-i\widetilde{\mathcal{H}}_+}\approx 1-i\widetilde{\mathcal{H}}_+-\frac{\widetilde{\mathcal{H}}_+^2}{2}
\approx e^{-iV}
 \begin{pmatrix}
   \gamma &\alpha \\
   -\alpha^* & \gamma^*
 \end{pmatrix},
\end{equation}
where 
\begin{eqnarray}
\gamma(x,y) &=& \cos\delta\cos\left(-i\partial_y-A_y\right)e^{-i\left(-i\partial_x-A_x\right)},
\nonumber
\\
\alpha(x,y) &=& e^{i\left(-i\partial_y-A_y\right)}\left[\sin\delta+i\sin\left(-i\partial_y-A_y\right)\right].
\nonumber
\end{eqnarray}
Here, $\gamma^*$, $\alpha^*$ are the corresponding complex conjugates.

\begin{figure}[b]
\begin{center}
\includegraphics[width=3.3in]{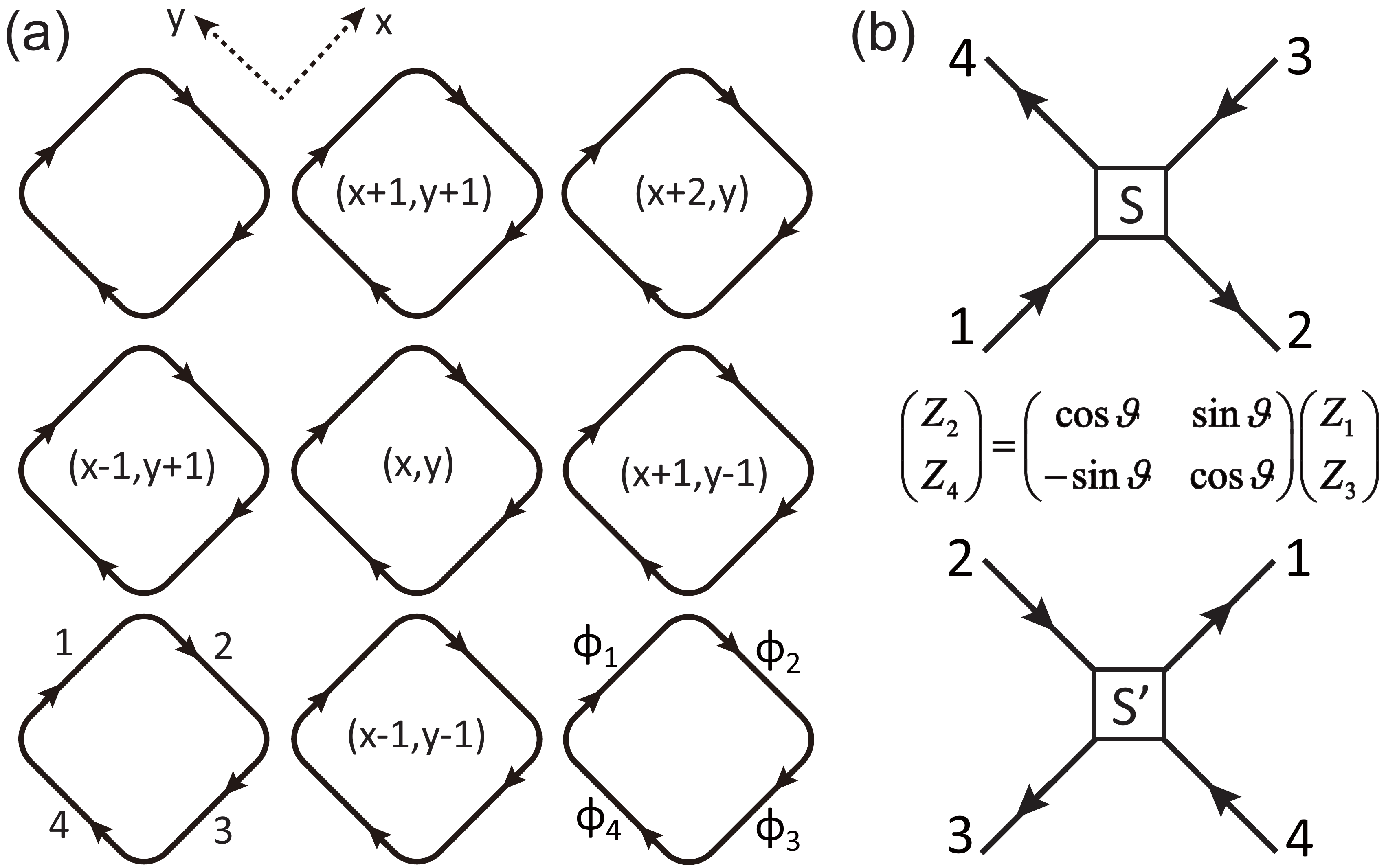}
\end{center}
\caption{The network model. (a) shows the coordinate system for plaquettes and the labeling of the four links.
(b) Amplitudes associated with possible scattering paths at nodes.}
\label{fig2}
\end{figure}

Then, we turn to the network model as shown in Fig.~\ref{fig2}. Such model is defined using the language of scattering theory~\cite{chalker1988}.
It consists of a square lattice of plaquettes. At the boundary of each plaquette, there is an edge state at the Fermi energy representing equipotentials, in which an electron drifts along the direction indicated by the arrow. Plaquettes are labeled by integer coordinates $(x,y)$, and we denote the four links $i$ making up a plaquette by $i=1, 2, 3, 4$, so that a link is specified by the combination $(x,y,i)$ where $x+y$ is even. The wave function for the electron on the link $(x,y,i)$ is represented by the current amplitude $Z_i(x,y)$, which is characterized by the phase change $\phi_i$ along the link ($0\leq\phi_i\leq2\pi$). The tunneling process at the nodes [denoted as $\mathbf{S}$ and $\mathbf{S}'$ in Fig.~\ref{fig2}(b)] may be related by a scattering matrix with a parameter $\vartheta$ ($0\leq\vartheta\leq\pi/2$) as
\begin{equation}
\begin{pmatrix}
Z_2\\
Z_4
\end{pmatrix}
=\begin{pmatrix}
\cos\vartheta & \sin\vartheta\\
-\sin\vartheta & \cos\vartheta
\end{pmatrix}
\begin{pmatrix}
Z_1\\
Z_3
\end{pmatrix}.
\end{equation}
Now, we associate a unitary scattering matrix with the model~\cite{ho1996}, which is roughly a time evolution operator. In the basis of $\left(Z_1(x,y), Z_3(x,y); Z_2(x,y), Z_4(x,y)\right)$, the one-step scattering matrix between the nearest-neighbor links is
\begin{equation}
\mathcal{S} = \begin{pmatrix}
0 & \mathcal{N}_1 \\
\mathcal{N}_2 & 0
\end{pmatrix},
\end{equation}
where
\begin{equation}
\mathcal{N}_1 = \begin{pmatrix}
\sin\vartheta e^{i\phi_1}\tau^x_-\tau^y_+ & \cos\vartheta e^{i\phi_1} \\
\cos\vartheta e^{i\phi_3} & -\sin\vartheta e^{i\phi_3}t^x_+t^y_-
\end{pmatrix},
\nonumber
\end{equation}
and
\begin{equation}
\mathcal{N}_2 = \begin{pmatrix}
\cos\vartheta e^{i\phi_2} & \sin\vartheta e^{i\phi_2}\tau^x_+\tau^y_+ \\
\sin\vartheta e^{i\phi_4}\tau^x_-\tau^y_- & -\cos\vartheta e^{i\phi_4}
\end{pmatrix}.
\nonumber
\end{equation}
Here, $\tau^x_\pm$ and $\tau^y_\pm$ are the translation operators defined as $\tau^x_{\pm}Z_i(x,y)=Z_i(x\pm1,y)$ and $\tau^y_{\pm}Z_i(x,y)=Z_i(x,y\pm1)$. The two-step scattering matrix then decouples as
\begin{equation}
\mathcal{S}^2 = \begin{pmatrix}
\mathcal{N}_1\mathcal{N}_2 & 0 \\
0 & \mathcal{N}_2\mathcal{N}_1
\end{pmatrix}.
\end{equation}
To extract the localization length, it is sufficient to just deal with the upper-left block $\mathcal{N}_1\mathcal{N}_2$~\cite{ho1996}.
If the phases $\phi_i$ are uniformly distributed between $0$ and $2\pi$, the network model is critical at $\vartheta=\vartheta_c=\pi/4$ where $\xi$ diverges, and in the localized phase otherwise~\cite{chalker1988}.

In the continuum limit, the translation operators can be written as $\tau^{x}_\pm=e^{\pm\partial_{x}}$ and $\tau^{y}_\pm=e^{\pm\partial_{y}}$. By identifying $A_x=(\phi_1-\phi_3)/2$, $A_y=(\phi_4-\phi_2)/2$, $V=-\sum_{i=1}^{4}\phi_i/2$ and $\vartheta=\vartheta_c+\delta/2$, we find that the unitary matrix $\mathcal{N}_1\mathcal{N}_2$ is exactly the same as the evolution operator $\mathcal{U}$ defined in Eq.~(\ref{evolution}). Specifically, the randomness in the individual link phases arise from fluctuation in the vector potential $\vec{A}$, variations in the total Aharonov-Bohm phase associated with each plaquette come from fluctuations in the scalar potential $V$, and the random tunneling parameter is not constant everywhere if the fluctuations in the mass $\Delta$ are present. Similar procedure can be done for $\mathcal{H}_-(k)$ for $C=-1$ to $C=0$ transition. Therefore, by using of the time evolution operator, we have established in detail a mapping from the QAH plateau transition to the network model.

\section{Results and Discussion}
\subsection{Coercivity transition}

\begin{figure}[t]
\begin{center}
\includegraphics[width=3.4in]{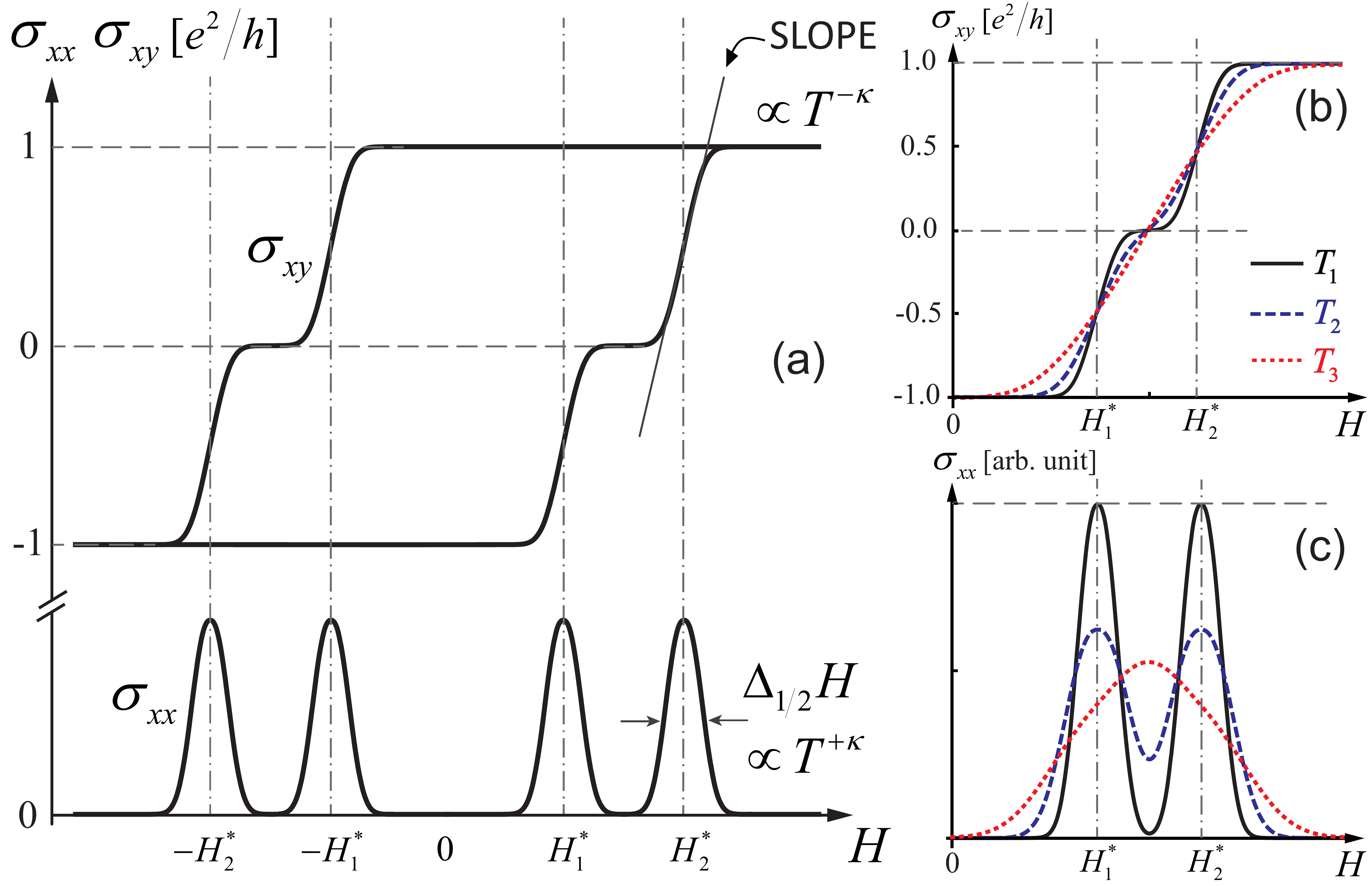}
\end{center}
\caption{(color online) Magnetic field dependence of $\sigma_{xy}$ and $\sigma_{xx}$. (a) Sketch of $\sigma_{xy}$ and $\sigma_{xx}$ as a function of applied magnetic field $H$. An intermediate plateau with $\sigma_{xy}=0$ appears at the hysteresis loop, while $\sigma_{xx}$ shows two peaks around the coercive field. (b) $\sigma_{xy}$ vs. $H$ at three different $T$ with $T_1<T_2<T_3$. (c) The corresponding $\sigma_{xx}$ vs. $H$.}
\label{fig3}
\end{figure}

The QAH plateau transition at the coercivity should have the same critical behavior as the network model. More specifically, the localization length $\xi$ of the levels near the Fermi energy diverges like a universal power law in $\Delta$ as $\xi = \xi_{\Delta}\left|\Delta-\Delta^*\right|^{-\nu}$. For $\Delta\propto M$, and at the coercivity $M\propto H$, therefore, 
\begin{equation}
\xi(H)=\xi_0\left|H-H^*\right|^{-\nu},
\end{equation}
with the critical exponent $\nu\approx2.4$ and $H^*$ is the critical external field of the plateau transition.
As there exist two critical points at $\Delta_1^*$ and $\Delta_2^*$, we predict there should be \emph{four} critical magnetic field $\pm H_1^*$ and $\pm H_2^*$ at which $\xi$ diverges as shown in Fig.~\ref{fig3}.

In the finite-size scaling theory, the conductance tensor depends on the parameter $H$ only through a single variable with the ansatz~\cite{pruisken1988},
\begin{equation}\label{scaling}
\sigma_{\alpha\beta}(H) = f_{\alpha\beta}\big[L^{1/\nu}_{\mathrm{eff}}\left(H-H^*\right)\big],
\end{equation}
where $\alpha,\beta=x,y$. $\sigma_{xx}$ is the longitudinal conductance. $L_{\mathrm{eff}}$ is the effective system size.
$f_{\alpha\beta}$ is a regular function (power series) of its argument except near the QAH plateaus.
Such power-law behavior of the transport coefficients reflects the two-parameter scaling of the conductance tensor~\cite{huckestein1995,pruisken1988}. When $L_{\text{eff}}\gg\xi$, one expect $f_{xx}\propto\exp(-L_{\text{eff}}/\xi)$.

At $T=0$~K, $L_{\mathrm{eff}}$ is equal to the system size $L$. At finite $T$, $L_{\mathrm{eff}}$ is given by the phase coherence length $L_{\mathrm{in}}$~\cite{thouless1977}, which behaves as $L_{\mathrm{in}}(T)\propto T^{-p/2}$ as $T\rightarrow0$~\cite{abrahams1981}. Then $L_{\text{eff}}^{1/\nu}\propto T^{-\kappa}$ with $\kappa=p/2\nu$. The $n$th derivative of the conductance tensor at the critical point is
\begin{equation}\label{Tscaling}
\frac{\partial^n\sigma_{\alpha\beta}(H^*)}{\partial H^n} \propto L_{\text{eff}}^{n/\nu} \propto T^{-n\kappa}.
\end{equation}
This is the $T$-dependent scaling of QAH plateau transition. More specifically, as shown in Fig.~\ref{fig3}(a), the maximum slope in the $\sigma_{xy}$ curve diverges as a power law in temperature $T$ as 
\begin{equation}
(\partial\sigma_{xy}/\partial H)_{\text{max}} \propto T^{-\kappa}.
\end{equation} 
In addition, the half-width of $\sigma_{xx}$ peak vanishes like 
\begin{equation}
\Delta_{1/2}H \propto T^{\kappa}. 
\end{equation}
The statement of Eq.~(\ref{Tscaling}) can be directly translated into resistance [see Appendix~\ref{res_con_tensor}].

The exponent $\nu$ can be measured directly by studying same Hall-bar geometries but different sizes. For sufficiently small samples, $(\partial\sigma_{xy}/\partial H)_{\text{max}}$ and $\Delta_{1/2}H$ should saturate at low $T$, and the saturation temperature would decrease with increasing system size. This is because that as the temperature when $L_{\text{in}}\sim L$, the $T$-dependent scaling at higher $T$ crosses over to size-dependent scaling. The saturation value of $\Delta_{1/2}H$ at low $T$ is then given by the condition $L/\xi\approx 1$, i.e., 
\begin{equation}
\Delta_{1/2}H \propto L^{-1/\nu}.
\end{equation}

The universal power-law behavior in temperature shows the characteristics of a second-order phase transition. And the magnetization $M$ is used as a continuous parameter for the phase transition between adjacent QAH phases. One may be concerned with this assumption, since in a FM material, $M$ is usually thought to reverse abruptly (known as the ``infinite avalanche'') at the coercivity, marking the occurrence of a first-order transition~\cite{vladimir1983}. Such discontinuity will completely conceal the above second-order phase transition. However, as studied extensively by materials scientists, the hysteresis curve of FM materials are often smooth. This is due to inevitable dissipations (such as the presence of disorders) in the process of magnetization~\cite{sethna1993}. The existence of dissipations make the magnetization process no longer a first-order transition, but a smooth crossover. Therefore, one could observe the critical behavior of QAH plateau transition on the hysteresis loop in a magnetic TI.

\subsection{Experimental proposal}

For the recent QAH experiment in a Cr$_x$(Bi,Sb)$_{2-x}$Te$_3$ thin film, at low enough $T$, one would observe the zero Hall plateau with $\rho_{xy}=0$ and $\sigma_{xy}=0$. The corresponding $\sigma_{xx}$ would have two peaks at the coercivity as shown in Fig.~\ref{fig3}(a), while $\rho_{xx}$ only has one peak. This remarkable theoretical prediction is already borne out in experiment~\cite{chang2013b}, by inverting the experimental data of $\rho_{xx}$ into $\sigma_{xx}$, $\sigma_{xx}$ shows double peaks at the coercivity while $\rho_{xx}$ only has single peak~\cite{ke_note}. However, the $\rho_{xy}=0$ and $\sigma_{xy}=0$ plateau are not yet observed, possibly because $T$ is still not low enough or the transitions in $\mathcal{H}_+(k)$ and $\mathcal{H}_-(k)$ are nearly degenerate~\cite{reason_note}.
As shown in Fig.~\ref{fig3}(b), the $\sigma_{xy}=0$ plateau disappears as $T$ increases.

Even without the signature of zero Hall plateau in $\rho_{xy}$, one can still measure the critical behavior by studying the $T$-dependent and size-dependent scaling predicted above. For a definite system size, the maximum slope in $\rho_{xy}$ should diverge in $T$ as
\begin{equation}
(\partial\rho_{xy}/\partial H)_{\text{max}} \propto T^{-\kappa}.
\end{equation}
However, the temperature dependence of the Fermi-Dirac distribution leads to a temperature dependence of the resistance,
$\rho_{\alpha\beta}(T)=\int dE (-\partial f(T)/\partial E)\rho_{\alpha\beta}(T=0)$. In order to observe the universal scaling
behavior, the temperature must be low enough that the influences of the finite width of Fermi-Dirac distribution can be neglected. 
While for a definite low temperature, the maximum slope in $\rho_{xy}$ scales in $L$ as 
\begin{equation}
(\partial\rho_{xy}/\partial H)_{\text{max}} \propto L^{-1/\nu}. 
\end{equation}
Moreover, $\rho_{xx}\propto\exp\left(-L_{\text{eff}}\left|H-H^*\right|^\nu/\xi_0\right)$ when $\rho_{xy}$ is close to the quantized value with $L_{\text{eff}}\gg\xi$. The critical exponent $\nu\approx2.4$, independent of the transition is degenerate or not~\cite{li2009,koch1991b,engel1993,chalker1988,lee1993,slevin2009}.

\section{Conclusion}

In summary, starting from the microscopic model for QAH plateau transition, we construct a mapping to the network model for integer QHE transition. We predict that $\sigma_{xx}$ would show two peaks at the coercivity while $\rho_{xx}$ only has single peak. Remarkably, this theoretical prediction is already borne out in experiment~\cite{chang2013b}. The scaling theory of Hall plateau transition in QAH effect is proposed. To observe the universal scaling behavior, $T$ must be low enough. However, the absolute scale in $T$ is very much dependent on the microscopic details of the randomness in magnetic domains. Only the value of the exponent $\nu$ is universal~\cite{kappa_note}. Moreover, without LLs, QAH plateau transition at the coercivity in a magnetic TI provides an experimental platform to test the random Dirac model~\cite{ludwig1994}, which was originally proposed for the description of integer QHE plateau transition. A field theory description of the QAH transition including the renormalization group flow of $\sigma_{xx}$ and $\sigma_{xy}$ will be studied in future work.

\begin{acknowledgments}
We are deeply indebted to Steven Kivelson for many valuable discussions. This work is supported
by the Defense Advanced Research Projects Agency Microsystems
Technology Office, MesoDynamic Architecture Program (MESO) through the Contract No.~N66001-11-1-4105; the DARPA Program on
``Topological Insulators -- Solid State Chemistry, New Materials and Properties,'' under the Grant No.~N66001-12-1-4034; and by the US Department of
Energy, Office of Basic Energy Sciences, Division of Materials Sciences and Engineering, under Contract No.~DE-AC02-76SF00515.
\end{acknowledgments}

\begin{appendix}

\section{Plateau transition point}
\label{plateautransition}

\begin{figure*}[t]
\begin{center}
\includegraphics[width=6.7in]{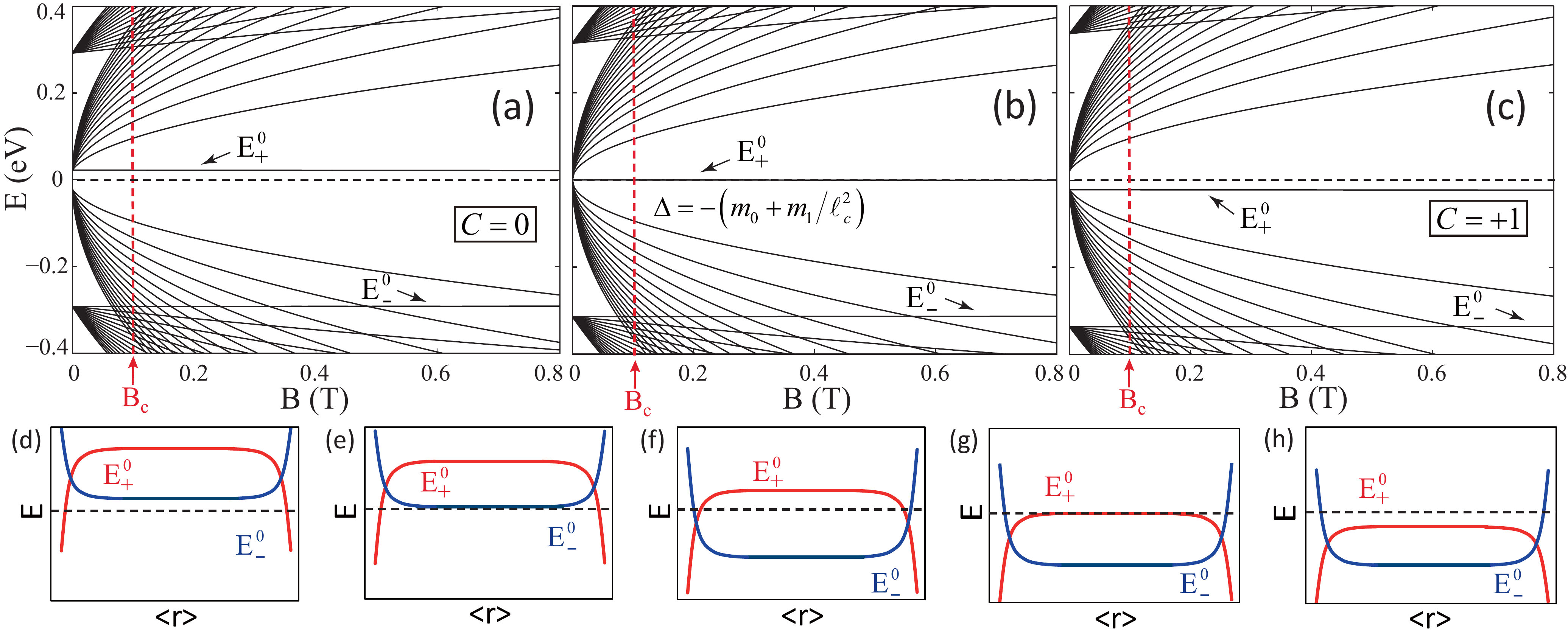}
\end{center}
\caption{The bulk and edge state spectrum of the QAH model described by Eq.~(1) in the presence of external magnetic field. (a), (b) \& (c) shows the bulk LLs, where in (a) $m_0+m_1/\ell_c^2<\Delta<-m_0-m_1/\ell_c^2$, in (b) $\Delta=-m_0-m_1/\ell_c^2$, and in (c) $\Delta>-m_0-m_1/\ell_c^2$. The Chern number (a) $C=0$, (b) transition point, (c) $C=+1$. The coercivity $B_c\approx0.097$~T, in (a)-(c) it clearly shows $|m_1/\ell_c^2|\ll |m_0|$. (d)-(h) shows the low-lying bulk and edge state energies as a function of the centers of the Landau orbitals when varying $\Delta$. $\Delta$ and corrosponding Chern number, (d) $\Delta<m_0+m_1/\ell_c^2$ and $C=-1$, (e) transition point $\Delta=m_0+m_1/\ell_c^2$, (f) $m_0+m_1\ell_c^2<\Delta<-m_0-m_1/\ell_c^2$ and $C=0$, (g) transition point $\Delta=-m_0-m_1/\ell_c^2$, (h) $\Delta>-m_0-m_1/\ell_c^2$ and $C=+1$. In (f),
the Fermi energy lies in-between the two bulk inverted LLs. The Fermi energy crosses the LLs, giving rise to the pair of counterpropagating edge states. It is the case for (a). (g) corresponds to (b). (h) corresponds to (c), where the Fermi energy only cross one LL, give rise to $C=1$.}
\label{fig4}
\end{figure*}

An external magnetic field is required to induce the coercivity transition, and in experiment the coercive field is small ($B_c<0.1$~T)~\cite{chang2013b}. There is no Landau levels (LLs) in this system as the cyclotron energy at the coercivity is much smaller than the potential fluctuation. Such small coercivity will shift the plateau transition point from $\Delta=\pm m_0$ to $\Delta=\pm (m_0+m_1/\ell_c^2)$, where $\ell_c=\sqrt{\hbar/eB_c}$ is the magnetic length. With $B_c<0.1$~T, $m_1/\ell_c^2<0.1$~meV. Since for the magnetic TI thin film studied in experiment $m_0\gg 1$~meV~\cite{chang2013b,wang2013b}, the shift of plateau transition point due to the coercivity is \emph{negligible}. This can be obtained by including the magnetic field in the Hamiltonian $\mathcal{H}_0$, and study the Chern number change as $\Delta$ varies.

At the coercivity, the external magnetic field enters into Eq.~(\ref{model1}) via minimal coupling: $\vec{k}\rightarrow \vec{k}+e\vec{A}$, where in the symmetric gauge the vector potential
\begin{equation}
\vec{A}=\frac{B}{2}\left(-y,x\right).
\end{equation}
We define the new operators
\begin{eqnarray}
\pi_{+} &=& \hbar\left(k_{+}+\frac{ieB}{2\hbar}z\right),\\
\pi_{-} &=& \hbar\left(k_{-}+\frac{ieB}{2\hbar}z^*\right),
\end{eqnarray}
where $k_\pm=k_x\pm ik_y$ and $z=x\pm iy$. These operators obey the commutation relations
\begin{equation}
\left[\pi_+,\pi_-\right]=-\frac{2\hbar^2}{\ell_c^2}.
\end{equation}
with the magnetic length $\ell_c=\sqrt{\hbar/eB}$. Using these commutation relation we define rasing and lowering operators
\begin{eqnarray}
&&a = \frac{\ell_c}{\sqrt{2}}\pi_-, \ \ a^{\dag} = \frac{\ell_c}{\sqrt{2}}\pi_+,
\\
&&\left[a, a^{\dag}\right] = 1.
\end{eqnarray}
The Hamiltonian can be rewritten as
\begin{eqnarray}
\mathcal{H}_0 &=& \begin{pmatrix}
\mathcal{H}_+(a,a^{\dag}) & 0\\
0 & \mathcal{H}_-(a,a^{\dag})
\end{pmatrix},
\\
\mathcal{H}_\pm(a,a^{\dag}) &=& \left[m_0\pm\Delta+\frac{2m_1}{\ell_c^2}\left(a^{\dag}a+\frac{1}{2}\right)\right]\tau_3
\nonumber
\\
&&+\frac{\sqrt{2}v_F}{\ell_c}\left(ia\tau_\pm-ia^{\dag}\tau_\mp\right).
\end{eqnarray}
where $\tau_j$ ($j=1,2,3$) are Pauli matrices, $\tau_\pm=(\tau_1\pm i\tau_2)/2$.

The spectrum of this Hamiltonian can be solved since only a finite number of harmonic oscillator Landau levels are coupled. The energy spectrum is
\begin{equation}
E_s = -s\frac{m_1}{\ell_c^2}\pm\sqrt{\frac{2v_F^2}{\ell_c^2}\mathcal{N}+\left(m_0+s\Delta+\frac{2m_1}{\ell_c^2}\mathcal{N}\right)^2}
\end{equation}
with $s=\pm$, and $\mathcal{N}=0,1,2,3,...$. This spectrum has ``zero mode'' given by
\begin{eqnarray}
E^0_+ &=& -m_0-\Delta-\frac{m_1}{\ell_c^2},\\
E^0_- &=& m_0-\Delta+\frac{m_1}{\ell_c^2}.
\end{eqnarray}
At the coercivity $B_c$, $E^0_{\pm}=0$ gives the transition point. Thus at half filling, the total Chern number of the system with the magnetic field becomes
\begin{equation}\label{chern}
C = \begin{cases}
\Delta/|\Delta|, & \text{for} \ \ \left|\Delta\right|>\left|m_0+m_1/\ell^2\right|\\
0, & \text{for} \ \ \left|\Delta\right|<\left|m_0+m_1/\ell^2\right|
\end{cases}
\end{equation}
Now the plateau transition point becomes $\Delta=\pm(m_0+m_1/\ell_c^2)$ with $B=B_c$.

The bulk LL and edge state spectrum for 5-quintuple layers (QLs) of Cr$_x$(Bi,Sb)$_{2-x}$Te$_3$ magnetic TI with different $\Delta$ is shown in Fig.~\ref{fig4}. The parameters are taken from Ref.~\onlinecite{wang2013b}, where $m_0<0$ and $m_1>0$. Fig.~\ref{fig4}(a) shows bulk LLs with $m_0+m_1/\ell_c^2<\Delta<-m_0-m_1/\ell_c^2$, in Fig.~\ref{fig4}(f) it shows the corresponding edge states, and there should be counter-propagating edge states that carry opposite Hall current. In Fig.~\ref{fig4}(c) and (h) with $\Delta>-m_0-m_1/\ell_c^2$, the LL spectrum changes where the Fermi energy is slightly above the two zero modes, and only one of them will provide edge state, which gives $C=1$.

\section{Random perturbations}
\label{randomperturbation}

Now we consider the random perturbations to the pure Hamiltonian of Eq.~(\ref{model0}). First, at the coercivity, the magnetic domains are being switched from up to down \emph{randomly}. The exchange field induced by local spin $\langle S\rangle$ in such random magnetic domains will give rise to
\begin{equation}
\widetilde{H}_{\Delta} = \Delta_z\widetilde{\sigma}_3\otimes1+\Delta_x\widetilde{\sigma}_1\otimes1+\Delta_y\widetilde{\sigma}_2\otimes1,
\end{equation}
where $\Delta_z$ is the exchange field along $z$ axis, $\Delta_{x,y}$ are the exchange field in the $x$-$y$ plane.

Second, top and bottom surface state will feel different random scalar potentials $V_1$ and $V_2$, respectively,
\begin{equation}
\widetilde{H}_{V} = \overline{V} 1\otimes1 +\delta V1\otimes\widetilde{\tau}_3,
\end{equation}
with $\overline{V}=(V_1+V_2)/2$, and $\delta V=(V_1-V_2)/2$.

Third, a small external magnetizing field $H$ is required to induce the coercivity transition. At the coercivity, the magnetization of the system $M$ is spatially random. So the magnetic field $B=\mu_0(H+M)$ in this system is also random in space, which couples to the system through a random vector potential $\vec{A}=(A_x,A_y)$, with the minimal coupling $\vec{k}\rightarrow\vec{k}-\vec{A}$, we have
\begin{equation}
\widetilde{H}_{A} = -A_y\widetilde{\sigma}_1\otimes\widetilde{\tau}_3+A_x\widetilde{\sigma}_2\otimes\widetilde{\tau}_3.
\end{equation}
All three types of randomness have been taken into account.

Then we make a unitary transformation to the basis of $|+\uparrow\rangle$, $|-\downarrow\rangle$, $|+\downarrow\rangle$, $|-\uparrow\rangle$ with $|\pm\uparrow\rangle=(|t\uparrow\rangle\pm|b\uparrow\rangle)/\sqrt{2}$
and $|\pm\downarrow\rangle=(|t\downarrow\rangle\pm|b\downarrow\rangle)/\sqrt{2}$. The pure Hamiltonian decouples as
\begin{eqnarray}
\mathcal{H}_0(k_x,k_y) &=&
\begin{pmatrix}
\mathcal{H}_+(k) & 0\\
0 & \mathcal{H}_-(k)
\end{pmatrix},
\\
\mathcal{H}_{\pm}(k) &=& k_y\tau_1\mp k_x\tau_2+\left(m(k)\pm\Delta\right)\tau_3.
\end{eqnarray}
$\tau_i$ are Pauli matrices. The random perturbations in the new basis are
\begin{equation}
\mathcal{H}_{\Delta} = \begin{pmatrix}
\Delta_z\tau_3 & \Delta_x1_{2\times2}-i\Delta_y\tau_3\\
\Delta_x1_{2\times2}+i\Delta_y\tau_3 & -\Delta_z\tau_3
\end{pmatrix},
\end{equation}
\begin{equation}
\mathcal{H}_{V} = \begin{pmatrix}
\overline{V} & \delta V\tau_1\\
\delta V\tau_1 & \overline{V}
\end{pmatrix},
\end{equation}
\begin{equation}
\mathcal{H}_{A} = \begin{pmatrix}
-A_y\tau_1+A_x\tau_2 & 0\\
0 & -A_y\tau_1-A_x\tau_2
\end{pmatrix}.
\end{equation}

The $\Delta_{x,y}$ will in general mix $\mathcal{H}_+(k)$ and $\mathcal{H}_-(k)$. However, we only consider the plateau transition, and the transition point of $\mathcal{H}_+(k)$ and $\mathcal{H}_-(k)$ are different, $\Delta=-m_0$ for $\mathcal{H}_+(k)$ and $\Delta=m_0$ for $\mathcal{H}_-(k)$. For plateau transition at $\Delta=-m_0$, $\mathcal{H}_-(k)$ is gapped, and the low energy physics is only determined by $\mathcal{H}_+(k)$. The $\Delta_{x,y}$ term can be perturbatively added into $\mathcal{H}_+(k)$ as
\begin{equation}
\mathcal{H}_{\Delta}^{x,y} \approx \frac{\Delta_x^2+\Delta_y^2}{2\Delta}\tau_3,
\end{equation}
which gives a random exchange field along $z$ axis. In general, in the system the fluctuation $\Delta_{x,y}\ll\Delta$, thus $(\Delta_x^2+\Delta_y^2)/2\Delta\ll\Delta_z$. Therefore, to the first order, this term can be neglected. Similarly for the transition at $\Delta=m_0$.

$\delta V$ term will also mix $\mathcal{H}_+(k)$ and $\mathcal{H}_-(k)$. Following the same discussion above for $\Delta_{x,y}$. At $\Delta=-m_0$, $\delta V$ contributes a random exchange field term along $z$ axis in $\mathcal{H}_+(k)$ as
\begin{equation}
\mathcal{H}_{\delta V} \approx \frac{(\delta V)^2}{2m_0}\tau_3,
\end{equation}
$\delta V\ll |m_0|$, so this term is negligibly small compared to $\Delta_z$. Besides, the 2D film of magnetic topological insulator is very thin (less than 5~nm), the random scalar potential feeled by the top and bottom surface states are almost the same $V_1\approx V_2$. Therefore, $\delta V$ can be ignored. Similarly for the transition at $\Delta=m_0$.

Finally, we have build up the model for the QAH plateau transition,
\begin{eqnarray}
\mathcal{H} =\mathcal{H}_0 + \mathcal{H}_{\Delta} +\mathcal{H}_{V}+\mathcal{H}_A,
\end{eqnarray}
where
\begin{eqnarray}
\mathcal{H}_{\Delta} &=& \begin{pmatrix}
\Delta_z\tau_3 & 0\\
0 & -\Delta_z\tau_3
\end{pmatrix},
\\
\mathcal{H}_{V} &=& \begin{pmatrix}
\overline{V} & 0\\
0 & \overline{V}
\end{pmatrix},
\\
\mathcal{H}_{A} &=& \begin{pmatrix}
-A_y\tau_1+A_x\tau_2 & 0\\
0 & -A_y\tau_1-A_x\tau_2
\end{pmatrix}.
\end{eqnarray}
Redefine $\Delta_z(x,y)=\Delta(x,y)$ and $\overline{V}(x,y)=V(x,y)$, this gives Eq.~(\ref{randompotential}) in the paper.

\section{Resistivity and conductivity tensor}
\label{res_con_tensor}

The resistivity tensor is
\begin{equation}
\boldsymbol\rho = \begin{pmatrix}
\rho_{xx} & \rho_{xy}\\
-\rho_{xy} & \rho_{yy}
\end{pmatrix},
\end{equation}
and the conductivity tensor is
\begin{equation}
\boldsymbol\sigma = \begin{pmatrix}
\sigma_{xx} & \sigma_{xy}\\
-\sigma_{xy} & \sigma_{yy}
\end{pmatrix},
\end{equation}
with $\boldsymbol{\sigma}=\boldsymbol{\rho}^{-1}$, we have
\begin{equation}
\rho_{xx} = \frac{\sigma_{xx}}{\sigma_{xx}\sigma_{yy}+\sigma_{xy}^2}=\frac{\sigma_{xx}}{\sigma_{xx}^2+\sigma_{xy}^2},
\end{equation}
and
\begin{equation}
\rho_{xy} = \frac{-\sigma_{xy}}{\sigma_{xx}\sigma_{yy}+\sigma_{xy}^2}=\frac{-\sigma_{xy}}{\sigma_{xx}^2+\sigma_{xy}^2}.
\end{equation}
This transforms $\sigma_{xx}$ and $\sigma_{xy}$ into $\rho_{xx}$ and $\rho_{xy}$.

When $\Delta<-|m_0|$, the system is insulating with Chern number $C=-1$, thus we have
\begin{equation}
\boldsymbol\sigma = \frac{e^2}{h}\begin{pmatrix}
0 & -1\\
1 & 0
\end{pmatrix},
\end{equation}
and corresponding resistivity tensor is
\begin{equation}
\boldsymbol\rho = \frac{h}{e^2}\begin{pmatrix}
0 & 1\\
-1 & 0
\end{pmatrix}.
\end{equation}
Similar case for $\Delta>|m_0|$. When $-|m_0|<\Delta<|m_0|$, the system is insulating with Chern number $C=0$, thus we expect the conductivity tensor
\begin{equation}
\boldsymbol\sigma = \begin{pmatrix}
\eta & 0\\
0 & \eta
\end{pmatrix},
\end{equation}
where in large sample at zero-temperature ($T=0$), $\eta\rightarrow 0^+$; for finite sample with finite $T$, $\eta$ is very small (possibly due to variable range hopping). Thus the corresponding resistivity tensor is
\begin{equation}
\boldsymbol\rho = \begin{pmatrix}
1/\eta & 0\\
0 & 1/\eta
\end{pmatrix}.
\end{equation}
For the QAH effect in magnetic TI, at low $T$, there should exist zero Hall plateau with $\sigma_{xy}=0$ and $\rho_{xy}=0$. From the scaling theory, we predict that $\sigma_{xx}$ generally become nonzero between the plateau transition from $\sigma_{xy}=-e^2/h$ to $\sigma_{xy}=0$ and $\sigma_{xy}=0$ to $\sigma_{xy}=e^2/h$. At $\sigma_{xy}=0$ plateau, $\sigma_{xx}\rightarrow0$. Therefore, $\sigma_{xx}$ shows two peaks at the coercivity. However, $\rho_{xx}$ only shows one peak at the coercivity. Because at $\rho_{xy}=0$ plateau, $\rho_{xx}=1/\eta\rightarrow\infty$. In fact, this remarkable theoretical prediction is already borne out in experiment, by inverting the experimental data of $\rho_{xx}$ into $\sigma_{xx}$, at the coercivity, $\sigma_{xx}$ shows double peaks with two critical fields while $\rho_{xx}$ only has single peak~\cite{chang2013b}. 

The critical field $H_{1}^*$ and $H_2^*$ is not universal. For example, a slightly macroscopic inhomogeneity in the electron density across the sample will in general result a slightly different $H_{1}^*$ and $H_2^*$. Such inhomogeneities do not affect the power-law behaviors in $\rho_{xx}$ and $\rho_{xy}$.

\end{appendix}


\begin{thebibliography}{38}
\bibitem [{rea()}]{reason_note}%
  \BibitemOpen
  \href@noop {} {}\bibinfo {note} {For the disordered sample with a rather low
  mobility in experiment ($<800$~cm$^2$/Vs), the disorder broadening of levels
  may exceed the gap between two critical points $\Delta=\pm m_0$, so that
  transitions in $\mathcal{H}_+(k)$ and $\mathcal{H}_-(k)$ become degenerate,
  and the Chern number changes by 2 at the critical point.}\BibitemShut {Stop}%
\bibitem [{kap()}]{kappa_note}%
  \BibitemOpen
  \href@noop {} {}\bibinfo {note} {The temperature exponent $\kappa$ may not be
  universal. For $\kappa=p/2\nu$, the inelastic scattering expoenent $p$
  depends on the disorder configuration, and is not universal.}\BibitemShut
  {Stop}%
\end{thebibliography}
\end{document}